\documentclass[aps,prb,twocolumn,floatfix]{revtex4}
\usepackage{graphicx}
\usepackage{nature}
\usepackage{amsmath}

\begin{document}

\title{ Microscopic origin of Magnetic Ferroelectrics in Nonlinear Multiferroics}
\author{Jiangping Hu}
\affiliation{Department of Physics, Purdue University, West
Lafayette, IN 47007}
\date{\today}
\begin{abstract}
 A simple but general microscopic mechanism to understand
the interplay between the electric and magnetic degrees of freedom
is developed. Within this mechanism, the magnetic structure
generates an electric current which induce an counterbalance
electric current from the spin orbital coupling. When the magnetic
structure is described by a single order parameter, the electric
polarization is determined by the single spin orbital coupling
parameter, and the material is predicted to be a half insulator.
This mechanism provides a simple estimation of the value of
ferroelectricity and sets a physical limitation as well.
\end{abstract}


\maketitle

Multiferroics are materials in which magnetic and electric orders
are strongly coupled, and have attracted increasing
attention\cite{Tokura2006,Cheong2007}. Recent experimental research
on multiferroics has shown that ferroelectricity and magnetism not
only coexist in the same material but also couple so strongly that
the magnetic degree of freedom can be manipulated by an electric
field and the electric degree of freedom can be manipulated by a
magnetic
field\cite{Kimura2003a,Hur2004a,Kigomiya2003,Hur2004b,Chapon2004a,Kadomtseva2006,Katsura2007}.
This property promises important technological applications in the
future. However, the strong coupling between these two degrees of
freedom poses enormous challenges to theoretical attempts to
understand the microscopic mechanism inside the materials.
 So far, due to the complexity of such
materials, theoretical understanding of the microscopic mechanism of
strong magnetoelectric coupling is quite limited.

Through symmetric analysis in the Ginzburg-Landau approach to
thermodynamics, M. Mostovoy\cite{Mostovoy2006} has shown that, the
relationship between the ferroelectric order $\vec P$ and magnetic
order $\vec M$ in spiral magnets  is given by
 \begin{eqnarray} \label{eq1}
\vec P\propto  \vec M \times ( \vec \nabla \times \vec M).
\end{eqnarray}
Eq.1 qualitatively explains experimental results. Microscopically, a
mechanism, an inverse effect of the Dzyaloshinskii-Moriya
(DM)\cite{Dzyaloshinskii1958, MORIYA1960} interaction, to generate
Eq.\ref{eq1} has been proposed\cite{Katsura2005}. The inverse DM
mechanism is based on the idea that spin currents are induced
between the noncollinear spins, and can therefore be considered as
electric moments. This is the only known mechanism to date. However,
this mechanism is limited in two important ways: (1) it lacks
quantitative prediction results; (2) strictly speaking, the concept
of spin current in spin orbital coupled system is not well defined.
This Letter proposes a simple but general new mechanism to
understand  Eq.1. Based on well defined conventional concept,
electric current, the proposed mechanism is completely different
from the inverse DM mechanism. The proposed mechanism not only can
provide quantitative prediction but also enable us to consider a
unified picture of strong magnetoelectric coupling in multiferroics.

We know from Maxwell equation that static magnetism and static
electricity do not couple with each other, and that dynamic
electricity such as an electric current can generate a
magnetization. The inverse process also exists, namely,  a
magnetization can generate an electric current.  Let's revisit the
definition of the electric current. The current operator of the
electron is defined as the change in Hamiltonian with respect to the
variation of   the vector potential of electromagnetic field,
 i.e.
\begin{eqnarray}\label{def}
\vec j = -c \frac{\delta H}{\delta\vec A}.
\end{eqnarray}
In non-relativistic quantum mechanics, the definition of the
electric current includes three terms generated from three different
physics: (1) the contribution of standard momentum; (2) the spin
contribution illustrated in standard quantum textbooks\cite{Landau};
(3) the contribution of  spin orbital coupling. To be more specific,
we consider a single electron in a band structure. The electron is
described by  the Hamiltonian
\begin{eqnarray}\label{hh}
 H_e =\frac{(\vec p-e\frac{\vec A}{c})^2}{2m^*}+\alpha (\vec p-e\frac{\vec A}{c})\cdot(\vec \sigma \times \vec \nabla V(\vec
 r))-\mu(\vec \nabla \times \vec A)\cdot \vec \sigma
\end{eqnarray}
where, $m^*$ is the effective mass of electrons, $\alpha$ is the
effective spin orbital coupling parameter, $\mu=\frac{ge}{2mc}$ and
$\vec \sigma$ is the spin of the electron. In the absence of the
external electrodynamic field, i.e. $\vec A=0$, for a given
wavefunction, $\Psi(\vec r)$, the electric current derived from
Eq.\ref{def}
 is given by
\begin{eqnarray}\label{current}
\vec j=  \vec j_0 +\mu c\nabla \times (\Psi^*\sigma\Psi) +\alpha e
(\Psi^*\sigma\Psi) \times \nabla V(\vec r)
\end{eqnarray}
 where \begin{eqnarray} j_0=\frac{ie\hbar}{2m^*}[(\nabla
\Psi^*)\Psi-\Psi^*(\nabla \Psi)].
\end{eqnarray}
The three terms in Eq.4 precisely correspond to the three physics
discussed above.  The first term  usually dominates over the other
two terms in transport. Therefore, the last two terms  are usually
ignored and are  not familiar to most people. Here we show that in
multiferroics which are insulators, the interplay between the last
two terms in Eq.\ref{current} provides a fundamental mechanism to
understand the magnetoelectric coupling.

In the nonlinear multiferroics, $RMnO_3$, a nonlinear spiral
magnetic order has been observed\cite{Kimura2003a}. The spiral
magnetic order is formed by the localized spins of Mn atoms. To
study the electronic physics, we can use a Kondo-lattice type of
model. We consider that the electrons in the band couple to the
localized spins of the Mn atoms through spin exchange coupling.
Through the exchange coupling, we can naturally assume that the
magnetic ordering of the localized spins also generates the same
magnetic ordering for the electrons in the band. Now consider that
the magnetization of  the electrons in the band is a simple spiral
magnetic ordering
\begin{eqnarray} \label{spiral} \vec M_0=M_0( cos
qx/a, sin qx/a,0).
\end{eqnarray}
The electric current associated with the magnetization is
given by
\begin{eqnarray}
\vec J_M=\mu c\nabla\times\vec M_0=\frac{\mu cqM_0}{a}(0,0,cosqx/a)
\end{eqnarray}
which is a current along z direction. In fact, it is a `global'
current along z direction for a fixed x coordinate.  In an
insulator, the net electric current with  such a configuration
 must be zero based on Kohn's proof of the insulating property\cite{Kohn1964}, namely, the
magnetization current must be counterbalanced by other electric
currents. The total electric current contributed from $\vec j_0$ in
the band  also vanishes since the lattice mirror symmetry in the x-y
plane is not broken  in the nonlinear
multiferroics\cite{Kimura2003a,Goto2004,Kimura2005} in the absence
of external magnetic field. Therefore, the electric current from the
magnetic ordering must be counterbalanced by the electric current
induced from the spin orbital coupling.

The above analysis can also apply to  general nonlinear magnetic ordering structures which  induce similar electric
currents.  From Eq.\ref{current}, the cancelation requirement leads to
\begin{eqnarray} \label{mm}
\mu c\nabla \times \vec M_0 +\alpha e \vec M_0 \times \nabla V(\vec
r)=0
\end{eqnarray}
 By simple
algebraic modification and averaging in the total space, Eq.\ref{mm}
beomes
\begin{eqnarray}
\frac{\alpha e^2}{\mu c}<\vec E>= <\frac{(\vec M_0 \cdot \vec
 E)\vec M_0}{M_0^2}>+ <\frac{\vec M_0 \times (\vec \nabla \times\vec
M_0)}{M_0^2} >
\end{eqnarray}
where $<...>$ takes the space average and   $\nabla V(\vec r)=-e\vec
E(\vec r)$. The first term in the right side of above equation
usually vanishes when taking the space average for a space
modulating spin density. We obtain the total ferroelectricity as
\begin{eqnarray}\label{final}
\vec P =  \frac{\epsilon_0 \mu c}{\alpha e^2}<\frac{\vec M_0 \times
(\vec \nabla \times\vec M_0)}{M_0^2} >
\end{eqnarray}

Eq.\ref{final} is consistent with   Eq.\ref{eq1}. However,
Eq.\ref{final} provides detailed coupling coefficients that   are
different from the results   normally expected from  the
Ginzburg-Landau theory. First, the coefficients are inversely
proportional to effective spin orbit coupling, which is against
intuitive expectation. Second, Eq.\ref{final} predicts that
saturated value of ferroelectricity does not depend on the amplitude
of the magnetization.  Therefore, it suggests that increasing large
magnetization will not dramatically increase ferroelectricity, which
is against that normally expected from the simple Ginzburg-Landau
Theory. Finally, in Eq.\ref{final}, there is only one free
parameter, the effective spin orbital coupling.  The fact of the
existence of only one free parameter makes relatively easier to test
the new mechanism experimentally.

There are two important issues regarding of the above results.
First, it is clear that the above counter-intuitive dependence of
the polarization on the spin-orbit coupling parameter and the
magnitude of magnetization can not be correct for arbitrary small
values of these parameters. Therefore, what is the limitation of
this mechanism based on the current cancelation?  To answer this
question, we have to compare the energy saved from the current
cancelation and the energy cost from the polarization.  When a
polarization is developed, there is an energy paid for the
deformation of  the lattice. This energy cost $E_{cost}$ for a small
polarization is expected to be
\begin{eqnarray}
E_{cost} = \frac{1}{2}\lambda \vec P^2
\end{eqnarray}
where $\lambda$ depends on the detailed lattice structures.  The
energy saved in the current cancelation for the electrons,
$E_{save}$  can also easily estimated,
\begin{eqnarray}
E_{save}=\frac{m^*\mu^2c^2}{2e^2}( \nabla \times\vec M_0)^2.
\end{eqnarray}
In order to favor the current cancelation mechanism, we must have
$E_{save}>E_{cost}$, which leads to the following criteria for the
mechanism of the current cancelation by plugging Eq.\ref{final},
\begin{eqnarray}\label{condition}
\alpha M_0 > \sqrt{\frac{\epsilon_0^2\lambda}{m^* e^2}}
\end{eqnarray}
This criteria is satisfied in the spiral magnets  such as $TbMnO_3$
as we will show later.

Second, a careful reader may notice that there is a critical flaw in
the above derivation for a simple band picture: for any magnetic
ordering  of the localized spins of atoms, the electrons in a fully
filled band do not have magnetization response regardless of the
strength of spin exchange coupling between them. Namely, $\vec
M_0=0$ for any magnetic order $\vec M$ of the localized spins of
atoms. Thus the cancelation of the current in Eq.\ref{mm} does not
exist. This observation leads to another important prediction of the
paper: in order to generate
 magnetoelectric coupling through the current cancelation, the multiferroics must be
 a `half' insulator. Let's consider a single spin degenerate band to understand the
 rational behind the prediction. The presence of   electron-electron
 interaction and  the exchange interaction
 between the electrons in the band and the localized spin moments can all cause a single spin degenerate band to
 split into several bands.
The simple picture in Fig.1 shows the splitting of the
 original bands in two. If both of the two new bands are fully
 occupied, no electric current will be generated by the magnetic
  ordering of the localized spins, because $\vec M_0 = 0$.
As a matter of fact, one can picture the physics as the
magnetization and electric currents are exactly the opposite in
direction in two new bands. However, if a gap exists between the two
new bands, and if
    the lower energy band is completely filled but the upper energy band is empty,
     which we call  it  as a  `half' insulator. The curl of the magnetic
     ordering generates real electric currents in the half insulator which
     needs to be
     counterbalanced by the electric current from the spin orbital coupling.

 Now we quantitatively discuss the
ferroelectricity predicted by  Eq.\ref{final}.  We rewrite
$\alpha=\frac{1}{2m\Delta_s}$ where $\Delta_s$ can be viewed as  the
effective spin split energy gap. For the spiral magnetic structure
in Eq.\ref{spiral}, if we choose reasonable parameters with  g
factor to be two, the lattice constant $a=1nm$ and $\Delta_s$ with
an unit of ev, we obtain
\begin{eqnarray}\label{num}
P_y =0.88 q\Delta_s [\mu C/(cm^2\cdot ev)]
\end{eqnarray}
If we make reasonable assumption that $\Delta_s$ varies from $0.1ev$
to $1ev $ and consider the fact that theoretical values are
generally larger than experimental measurements due to the existence
of disorder, we can conclude that values predicted by Eq.\ref{num}
are in the same order of experimental values observed in
experiments\cite{Goto2004,Kimura2005} for the spiral magnets such as
$TbMnO_3$ and $DyMnO_3$.

To show that the condition, Eq.\ref{condition}, is really satisfied
in the spiral magnets, we can estimate the cost energy $E_{cost}$
and the saving energy $E_{save}$.  Taking the polarization value
$P_y\sim0.1\mu C/cm^2$ observed in $TbMnO_3$\cite{Kimura2003a}, we
can estimate the effective lattice shift, $\delta y=P_yV_n/e^*$,
where $V_n$ is the volume of one unit cell and $e^*=N_{eff} e $ is
the effective charge in one unit cell. The cost energy
$E_{cost}=\frac{1}{2}\kappa \delta_y^2$ where $\kappa$ is determined
by lattice structures. For $TbMnO_3$, $\kappa \sim 1ev/{{\AA}}^2$
and $V_n=2.3\times 10^2{\AA}^3$, we have $E_{cost}\sim
(\frac{0.014}{N_{eff}})^2 (ev)$. Let's assume that the  number of
effective spin of  electrons in one unit cell $M_0 = S_{eff}
(\hbar)$, $m^*=b m$ (m electron mass) and g=2, we obtain,
$E_{save}\sim 0.35 b S^2_{eff} (ev)$ for $q\sim 2\pi\times 0.3$ in
$TbMnO_3$. Comparing $E_{cost}$ and $E_{save}$, we see that the
Eq.\ref{condition} can be satisfied for $S_{eff}>0.02$. It is
important to note that the $S_{eff}$ is not the total effective spin
of magnetized atoms measured in experiments and it is always less
than one since there is no contribution from completely filled
bands.

 Although the condition, Eq. \ref{condition}, posts a limitation on the current cancelation mechanism,
 Eq.\ref{num}  suggests that the value of ferroelectricity can still  grow
largely as $\Delta_s$ increases. However, in the electric current
cancelation mechanism, there is an additional limitation  on  the
value of the polarization. The ferroelectricity is limited by a
natural energy scale, the energy gap $\Delta_g$ in the `half'
insulator. The gap defines a length scale $l= \frac{h}{\sqrt{2m^*
\Delta_g}}$. In order to maintain the validity of a `half'
insulator, the value of ferroelectricity must satisfy the following
criteria,
\begin{eqnarray} \label{limit}
|P| <\frac{\epsilon_0 \Delta_g}{ e l}=0.72 \sqrt{b \Delta_g^3} [\mu
C/(cm^2\cdot ev^{\frac{3}{2}})]
\end{eqnarray}
 the energy unit of $\Delta_g$
is ev. Eq.\ref{limit} shows that it is very hard for the
ferroelectricity in the multiferroics to be larger than  a few $\mu
C/ cm^2$ because $\Delta_g$  should be in a range of a few ev at
most. This explains why the ferroelectricity measured in the
multiferroics
 is much lower than that  in  conventional ferroelectric
materials. For example, the largest ferroelectricity in the family
of $RMnO_3$, $R=Dy,Tb,Gd...$  is  $P\sim 0.3\mu
C/cm^2$\cite{Kimura2005} which is measured in $DyMnO_3$ while in
conventional ferroelectric materials, such as $BaTiO_3$,  $ P=26 \mu
C/cm^2$\cite{Lines}.

In summary, this letter develops a new mechanism to explain
magnetoelectric coupling.  With a
relative small number of free parameters, the mechanism can estimate
the value of ferroelectricity. The free parameters can be
independently measured using various experimental techniques. For
example, the energy gap in the multiferroics and the spin orbit
coupling strength can be measured in optical absorption spectrum.
The prediction of a `half' insulator can be tested via numerical
calculations as well. Most importantly, the mechanism sets a general
guideline for the search of new multiferroics materials with
ferroelectricity of larger value.

The new mechanism  also   provides a new perspective through which
magnetoelectric coupling can be understood:   focusing on electronic
physics rather than   on  lattice dynamics. The lattice dynamics has
been the key to understand conventional ferroelectricity materials
because
 the development of ferroelectricity is tied to phonon softening\cite{Lines}. However,
  in many multiferroics, no clear indication for the phonon softening has been found\cite{Cheong2007,Sirenko}.
  Our new mechanism clearly states that in the multiferroics, it is the electronic  properties  that are responsible for the magnetoelectric coupling.

It is natural to ask if the mechanism we discussed above is
universal to all multiferroics. Although it is not easy to test, we
believe that the answer is positive. The analysis in this letter
assume the presence of a single magnetic order in the material. With
a single magnetic order, the electric current associated to the magnetic
order  is easy to be defined.   In many multiferroics, the magnetic
structure are very complicated and can not be  described by a single
magnetic order parameter\cite{Cheong2007,Hu2007}. A careful
construction of electric current due to magnetization is required.
These physics await future investigations.

 \begin{figure}
\includegraphics[width=5cm]{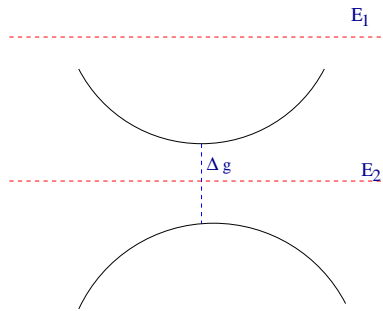}
\caption{\label{fig1}  A sketch of two bands which are split from
one spin degenerate band. $\Delta_g$ defines the band gap. When the
chemical potential at $E_1$, both bands are fully filled. When the
chemical potential at $E_2$, it is a `half' insulator which can
generate ferroelectricity.}
\end{figure}

The author thanks Prof. P. Muzikar and Prof. G. Giuliani for
extremely useful discussion. The author also thanks  Dr. A. Bernevig,  Dr. H.D Chen,  and Prof. S. C. Zhang for their useful comments and discussions.
The author also acknowledges  Fang Chen and
KangJun Seo for discussion.  This work was supported by the National
Science Foundation under grant number PHY-0603759.

\bibliography{ferroics}

\end{document}